\documentclass{elsart}
\usepackage{amssymb}\usepackage{bm}
\usepackage{amsmath}

\newcommand{\beq}{\begin{equation}}
\newcommand{\eeq}{\end{equation}}
\begin{document}
\begin{frontmatter}
\title{Variational Inequalities in Critical-State Problems
}
\author{Leonid Prigozhin}
\address{Department of Solar Energy and Environmental Physics\\
The Jacob Blaustein Institute
for Desert Research\\
Ben Gurion University of the Negev, Sede Boqer Campus, 84990
Israel} \ead{leonid@cs.bgu.ac.il}
\ead[url]{http://www.cs.bgu.ac.il/$\sim$leonid}
\begin{abstract}
Similar evolutionary variational inequalities appear as convenient
formulations for continuous quasistationary models for sandpile
growth, formation of a network of lakes and rivers, magnetization
of type-II superconductors, and elastoplastic deformations. We
outline the main steps of such models derivation and try to
clarify the origin of this similarity. New dual variational
formulations, analogous to mixed variational inequalities in
plasticity, are derived for sandpiles and superconductors.
\end{abstract}

\end{frontmatter}

\section*{Introduction} Spatially extended dissipative systems have recently
attracted much interest among physicists. These systems may have
infinitely many metastable states but, driven by the external
forces, often organize themselves into a marginally stable
``critical state'' and are then able to demonstrate almost
instantaneous long-range interactions between their distantly
separated parts.

Modifications of a crude cellular automata model of sandpile
\cite{BTW} have been used by many authors for simulating such
systems behavior and it was sometimes doubted whether the models
based on differential equations can in principle be employed:  the
relaxation in continuous models is expected to be a smooth process
evolving in time, while, e.g., cellular automata models are able
to mimic sand avalanches as sudden catastrophic events.
Nevertheless, continuous models allowing for long-range
interactions, hysteresis, metastability, and avalanches have been
derived for sandpiles \cite{PRE1,Evans1,Evans2}, river networks
\cite{PRE1}, type-II superconductors \cite{Boss,Euro2,IEEE}.
Although these are dissipative systems of a different nature,
their continuous models are equivalent to very similar variational
or quasivariational evolutionary inequalities, the formulations
convenient for both the numerical simulation \cite{GLT} and
theoretical study \cite{DL}. Much earlier, similar variational
formulations have been derived for various plasticity problems
(see \cite{DL}).

The aim of this work is to outline the main steps of such models
derivation and to clarify the origin of this similarity. We
present the simplest version of each model and try to avoid
mathematical details but give references to rigorous proofs,
numerical simulations, and possible extensions of the considered
models. We also derive new dual variational formulations in terms
of the conjugate variables for sandpiles and superconductors. The
dual problems are similar to mixed variational inequalities in
plasticity \cite{HR1}. Well-posedness of these new problems is yet
to be investigated.

\section{Variational formulations}
\subsection{Sandpiles} Let a cohesionless granular material,
characterized by its angle of repose $\alpha$, be poured out onto
a rough rigid surface $y=h_0(x)$, where $y$ is vertical and
$x\in\Omega\subset \Rset^2$. We find the shape of a growing pile,
$y=h(x,t)$.

Assuming the flow of granular material down the slope of the pile
is confined to a thin boundary layer and the bulk density of
material in the pile is constant, we can write the mass
conservation law in the form $\partial_t
h+\nabla\cdot\mbox{\boldmath$q$}=w,$ where $\mbox{\boldmath$q$}$
is the horizontal projection of the material flux and $w(x,t)\ge
0$ is the intensity of the source of material being poured onto
the pile. Neglecting inertia, we suppose that the surface flow is
directed towards the steepest descent,
$\mbox{\boldmath$q$}=-m\nabla h$, where \beq m(x,t)\ge 0
\label{m}\eeq is an {\it unknown} scalar function. The
conservation law assumes the form \beq\partial_t
h-\nabla\cdot(m\nabla h)=w.\label{eq}\eeq The free surface
initially coincides with the support, \beq
h(x,0)=h_0(x),\label{h0}\eeq and cannot lie below it, \beq
h(x,t)\ge h_0(x).\label{h>h0}\eeq Wherever the granular material
covers the support, the surface slope cannot exceed the material
 repose angle $\alpha$, \beq h(x,t)> h_0(x)\ \longrightarrow \
|\nabla h(x,t)|\le \gamma ,\label{h>}\eeq where
$\gamma=\tan(\alpha)$. No surface flow occurs over the parts of
the pile surface inclined less than at the angle of repose: \beq
|\nabla h(x,t)|< \gamma\ \longrightarrow \
m(x,t)=0.\label{eq_l}\eeq We assume for simplicity that there is a
vertical wall at the boundary $\Gamma$ of domain $\Omega$, hence
\beq m\partial_n h=0\  \mbox{on} \ \Gamma.\label{bound}\eeq The
formulated model for pile growth contains two unknowns, the free
surface $h$ and an auxiliary function $m$, and it is difficult to
deal with the equations and inequalities (\ref{m})-(\ref{bound})
directly. Fortunately, a more convenient variational formulation
can be derived (here we follow \cite{PRE1}, see \cite{Euro1} for a
rigorous proof).  Let us define, for every continuous function
$\psi$, a nonlinear operator
\[ B_{\psi}(\varphi)=\frac{1}{2}\left(|\nabla\varphi
|^2-M(\psi)\right), \] where \[ M(\psi)(x,t)=\left\{
\begin{array}{ll}
\gamma^2 & \mbox{if}\ \psi(x,t)>h_0(x),\\
\max(\gamma^2,|\nabla h_0(x)|^2)\ &\mbox{if}\ \psi(x,t)\le
h_0(x).\end{array}\right. \] We define also a family of closed
convex sets\footnote{More exactly, $ K(\psi)=\left\{ \varphi\in
L^{\infty}(0,T;W^{1,\infty}(\Omega)) \, |\, B_{\psi}(\varphi)\le
0\ \mbox{a.e.}\right\}$, see \cite{Euro1}.}
\[ K(\psi)=\left\{ \varphi(x,t) \ |\ B_{\psi}(\varphi)\le 0\
\right\},\] denote by $(u,v)$ the scalar product of two functions,
and consider an evolutionary
quasivariational inequality written for the pile surface alone:
\beq\begin{array}{c}\mbox{Find}\ h\in K(h)\ \mbox{such that}\
(\partial_t h -w,\varphi -h)\ge 0\ \mbox{for any}\ \varphi\in
K(h),\\
h(x,0)=h_0(x).\end{array}\label{qvi}\eeq

{\sc Theorem.} A function $h(x,t)$ is a solution of the
quasivariational inequality (\ref{qvi}) if and only if there
exists $m(x,t)$ such that the pair $\{h,m\}$ satisfies a weak form
of the problem (\ref{m})-(\ref{bound}).

{\sc Proof.} We formally rewrite the inequality (\ref{qvi}) as an
implicit optimization
problem \beq \begin{array}{ll} J_h(h)=&\min J_h(\varphi)\\
\ &\varphi\in K(h)\end{array}\label{opt}\eeq where
$J_h(\varphi)=(\partial_t h -w,\varphi)$ is a linear functional
which depends on the solution $h$. Let us fix the function $h$ in
$J_h$ and $K(h)$ and derive the necessary and sufficient condition
of optimality for (\ref{opt}) using the Lagrange multipliers
technique (\cite{ET}, ch. 3, th. 5.1). Substituting then the
function $h$ into this condition, we obtain a similar condition
for the problem with an implicit constraint: $h$ is a solution of
the quasivariational inequality (\ref{qvi}) if and only if there
exists a Lagrange multiplier $m(x,t)\ge 0$ such that the pair
$\{h,m\}$ is a saddle point of Lagrangian, i.e., \beq
J_h(h)+(m^{\prime},B_h(h))\le J_h(h)+(m,B_h(h))\le
J_h(h^{\prime})+(m,B_h(h^{\prime}))\label{saddle} \eeq for
arbitrary $h^{\prime}$ and $m^{\prime}\ge 0$. The condition of
supplementary slackness, \beq (m,B_h(h))=0,\label{slack}\eeq is
thereby fulfilled.\footnote{As is explained in \cite{Euro1}, to
satisfy a constraint qualification hypothesis (\cite{ET}, ch. 3,
(5.24)) we need to define $B_h:
L^{\infty}(0,T;W^{1,\infty}(\Omega))\rightarrow {\mathcal L}=L
^{\infty}((0,T)\times\Omega)$. Hence $m$ belongs to the dual space
${\mathcal{L}}^{\prime}$ and is a nonnegative Radon measure.}

Let $h$ be a solution of (\ref{qvi}). As follows from
(\ref{saddle}), the functional $ (\partial_t
h-w,h^{\prime})+\frac{1}{2}(m,|\nabla h^{\prime}|^2-M(h))$ has a
minimum at the point $h^{\prime}=h$. Hence, \beq (\partial_t
h-w,\chi)+(m,\nabla h\cdot\nabla \chi)=0\label{weak}\eeq for any
test function $\chi$. This is a weak formulation of equation
(\ref{eq}) with the boundary condition (\ref{bound}). Since $h\in
K(h)$, condition (\ref{h>}) is satisfied, (\ref{eq_l}) follows
from (\ref{slack}), and to show that $\{h,m\}$ satisfies all model
relations it remains only to check that $h\ge h_0$. Choosing
\[\varphi=\left\{\begin{array}{ll}
h+(h_0-h)_+\ &\mbox{for}\ 0\le t\le t_0,\\
h&\mbox{otherwise,}
\end{array}\right.\]
where $z_+$ means $\max(z,0)$ and taking into account that
$\varphi\in K(h)$, $w\ge 0$ we obtain \[0\le (\partial_t
h-w,\psi-h)\le
-\frac{1}{2}\int\left\{[h_0(x)-h(x,t_0)]_+\right\}^2d\Omega,\] so
the inequality (\ref{h>h0}) is proved.

Let now $\{h,m\}$ be a solution to (\ref{m})-(\ref{bound}). By
(\ref{h>h0}), (\ref{h>}) we have  $|\nabla h|\le \gamma$ wherever
$h>h_0$ and $h=h_0$ otherwise, hence $h\in K(h)$. To prove that
$h$ solves the quasivariational inequality, it is sufficient to
show the inequalities (\ref{saddle}) hold. It is easy to see that
(\ref{slack}) is true, so the left of the inequalities
(\ref{saddle}) is fulfilled. Using the weak form (\ref{weak}) of
equations (\ref{eq}), (\ref{bound}) we obtain, for
$\chi=h^{\prime}-h$,
\[J_h(h^{\prime})+(m,B_h(h^{\prime}))-J_h(h)-(m,B_h(h))=\frac{1}{2}(m,|\nabla
\{h^{\prime}-h\}|^2)\ge 0.\] Thus the second inequality in
(\ref{saddle}) holds too, which completes the proof.

We note that the auxiliary unknown, $m$, introduced into the pile
growth model to fix the possible  sand flux direction, turns out
to be a Lagrange multiplier related to the equilibrium constraint
upon the pile surface incline and is eliminated in transition to
the variational formulation. The multiplier depends in a non-local
way on the surface and source and that is why instantaneous long
range interactions over the critically inclined parts of the
surface are possible in this model. Such a situation is typical
also for other dissipative systems where the relaxation is fast
and the assumption that all the dynamics occur at the border of
stability is justified.

 If the support has no steep slopes,
$|\nabla h_0|\le \gamma$, the set of admissible functions $K$
becomes fixed (does not depend on the solution) and the inequality
(\ref{qvi}) becomes simply variational; in this case the existence
and uniqueness of a solution have been proved \cite{Euro1}. It
remains an interesting open problem to prove existence and
uniqueness in the general quasivariational case.

The variational formulation obtained is very convenient for
numerical simulation of pile growth, see \cite{ChESc}. There are
analytical solutions \cite{PRE1} exactly describing the pile
shapes in experiments \cite{Puhl}. Mathematically, the avalanches
upon pile surface correspond to solutions with the jumps caused by
sudden variations of the admissible functions set $K$ due to local
fluctuations of the repose angle, see \cite{PRE1}. Such
discontinuous solutions of the variational inequality have been
studied in \cite{Evans2}. It has been also shown \cite{PZ} that
the mesoscopic BCRE model \cite{BCRE} for sand surface dynamics
converges in the long scale limit to the inequality (\ref{qvi}).
In a continuous limit, stochastic cellular automata models of
sandpiles converge to a similar variational inequality with the
anisotropy inherited from the cellular structure of these models
\cite{ER}.

\subsection{Lakes and rivers} Let now $h_0$ be the land surface,
$w$ the intensity of precipitation. We assume for simplicity that
the water neither evaporates nor penetrates the soil but just
flows down the slopes and accumulates into lakes at local
depressions of the relief. The level of a lake rises until it
reaches the divide of two basins. Then a river running out of the
lake appears and transfers all additional water to another lake
below.

To model the evolution of this system of lakes and rivers, let us
note that the balance equation \[ \partial_t
h+\nabla\cdot\mbox{\boldmath$q$}=w,\label{qw}\] in which
$\mbox{\boldmath$q$}$ is the horizontal projection of water flux,
remains valid. The free boundary $h$ in this problem either
coincides with $h_0$ or, where it is higher, is the horizontal
surface of a lake: \[ h(x,t)\ge h_0(x),\ \ \
h(x,t)>h_0(x)\longrightarrow \nabla h(x,t)=0.\label{wcond}\] Over
the hill slopes, where $h=h_0$, we assume again that the flux is
in the steepest descent direction, \[
h(x,t)=h_0(x)\longrightarrow\mbox{\boldmath$q$}=-m\nabla h,\]
where $m(x,t)\ge 0$ is unknown. However, this is not true for the
lakes, where $h>h_0$ and $\nabla h =0$. In fact, although the lake
hydrodynamics are not trivial, the flow in the lake does not
affect the free surface, and it can be shown \cite{PRE1} that the
model relations above lead to the quasivariational inequality
(\ref{qvi}).

Indeed, let $\gamma =0$ and the pair $\{h,m\}$ satisfies these
relations. Then $h\in K(h)$ and for any function $\varphi\in K(h)$
we obtain
\[(\partial_t h-w,\varphi
-h)=-(\nabla\cdot\mbox{\boldmath$q$},\varphi
-h)=-\int^T_0\oint_{\Gamma}q_n(\varphi-h)+
\int^T_0\int_{\Omega}\mbox{\boldmath$q$}\cdot\nabla(\varphi-h).\]
The first integral on the right hand side is zero due to the
boundary conditions. Gradients $\nabla h$ and $\nabla\varphi$ are
both zero wherever $h>h_0$. Outside this domain $h=h_0$,
$\mbox{\boldmath$q$}=-m\nabla h_0$, $m\ge 0$, and $|\nabla \varphi
|\le |\nabla h_0|$. Therefore,
\[\mbox{\boldmath$q$}\cdot\nabla(\varphi-h)=-m(\nabla h_0\cdot\nabla\varphi
-|\nabla h_0|^2)\ge 0,\] the quasivariational inequality holds and
determines the free surface evolution.

This is, however, only a part of the solution needed: it is the
water flux $q=|\mbox{\boldmath$q$}|$, or, equivalently, the
auxiliary variable $m$, which is of interest in most
geomorphological and hydrological applications. Provided the free
surface $h$ is found, the water flux in the coincidence set
$h=h_0$ can be determined, at least in some simple cases
\cite{PRE1}, from the water balance equation which takes in this
set the form
\[-\nabla\cdot\left(q\frac{\nabla h_0}{|\nabla h_0|}\right)=w.\]
Generally, this is a difficult task and a different approach to
water flux calculation is desirable. Below, we consider an
alternative approach to determining the conjugate variables for
variational inequalities.

Note that if $\gamma=0$ the quasivariational inequality
(\ref{qvi}) is no more equivalent to the sandpile model
(\ref{m})-(\ref{bound}) in which ${q}=0$ if $\nabla h =0$. This
equivalency breaks down because, if $\gamma=0$, the constraint
qualification hypothesis  \cite{ET} is not true: there exists no
$\varphi$ such that $B_h(\varphi)<0$.

\subsection{Superconductors} Phenomenologically, the magnetic field
penetration into type-II superconductors can be understood as a
nonlinear eddy current problem. In accordance with the Faraday law
of electromagnetic induction, the eddy currents in a conductor are
driven by the electric fields induced by time variations of the
magnetic flux. Let the superconductor occupy a domain
$\Omega\subset\Rset^3$ and $\omega =
\Rset^3\backslash\overline{\Omega}$ be the outer space. We denote
by $\Gamma$ the common boundary of these domains and assume
$\bm{n}$, the unit normal to $\Gamma$, is directed outside
$\Omega$.

Omitting the displacement current in Maxwell equations and
assuming the magnetic permeability of the superconductor is equal
to that of vacuum and scaled to be unity, we obtain the following
eddy current model,
\beq\begin{array}{c}\partial_t\bm{h}+\nabla\times \bm{e}=0,\\
\nabla \times \bm{h}=\bm{j}+\bm{j}_e,\end{array}\ \ \ \ \ x\in
\mathbb{R}^3,\ t>0,\label{Mxwl}\eeq with the initial condition
$\bm{h}|_{t=0}=\bm{h}_0(x)$ such that $\nabla\cdot \bm{h}_0=0$.
Here $\bm{j}$ is the current in the superconductor ($\bm{j}=0$ in
$\omega$), and $\bm{j}_e$ is the given external current having a
bounded support $\mbox{\em supp}\{\bm{j}\}\subset \omega$ and
satisfying $\nabla\cdot \bm{j}_e=0$. Additionally, in the
conductive domain $\Omega$ a current-voltage law has to be
postulated.

In an ordinary conductor, the vectors of the electric field and
current density are related by the linear Ohm law. Type-II
superconductors are instead characterized in the Bean
critical-state model \cite{Bean} by a highly nonlinear
current-voltage relation which gives rise to a free boundary
problem. The problem has been solved mainly under the assumption
that the electric field and current density are parallel (see,
e.g., \cite{IEEE} and the references therein). Then
$\mbox{\boldmath$e$}=\rho\mbox{\boldmath$j$}$, where the {\it a
priory} unknown effective resistivity $\rho(x,t) \ge 0$
characterizes the energy losses accompanying movement of magnetic
vortices in a superconductor. It is assumed in the Bean model that
the current density $\mbox{\boldmath$j$}$ cannot exceed some
critical value $j_c$ and, until this value is reached, the
vortices are pinned and the current is purely superconductive:
\beq |\mbox{\boldmath$j$}(x,t)|\le j_c,\ \ \ \
|\mbox{\boldmath$j$}(x,t)|< j_c \longrightarrow
\bm{e}(x,t)=0.\label{BeanCond}\eeq

The simplest geometric configuration is that of a long
superconductive cylinder having a simply connected cross-section
$\Omega\subset \Rset^2$ and placed into a non-stationary parallel
uniform external magnetic field $\mbox{\boldmath$h$}_e(t)$. In
this case the most convenient variational formulation can be
derived for the magnetic field in the superconductor. The current
density $ \mbox{\boldmath$j$}$ induced by the external field
variations is parallel to the cross-section plane and produces the
magnetic field $\mbox{\boldmath$h$}_i(x,t)$, parallel to
$\mbox{\boldmath$h$}_e$ and equal to zero on $\Gamma$. The problem
is two-dimensional. Denoting by  $h_i$ and $h_e$ parallel to the
cylinder axis  components of $\bm{h}_i$ and $\bm{h}_e$,
correspondingly, and using the standard notations $curl\,
\mbox{\boldmath$v$}=\partial_{x_1}v_2-\partial_{x_2}v_1$,
$\mbox{\boldmath$curl$}\,v =(\partial_{x_2}v,-\partial_{x_1}v)$,
we rewrite the model (\ref{Mxwl}) as \beq\partial_t
(h_i+h_e)+curl\,\mbox{\boldmath$e$}=0,\ \ \
\mbox{\boldmath$curl$}\,h_i=\mbox{\boldmath$j$},\ \ \ x\in\Omega,\
t>0.\label{ME}\eeq Since $|\nabla h_i|=
|\mbox{\boldmath$curl$}\,h_i|=|\mbox{\boldmath$j$}|\le j_c$,
$h_i(x,t)$ should belong to the set \beq K=\{\varphi(x,t) \ |\
|\nabla\varphi |\le j_c,\ \varphi|_{\Gamma}=0\}.\label{KK}\eeq
Multiplying the first of equations (\ref{ME}) by $\varphi-h_i$,
$\varphi\in K$, integrating, and using the Green formula we obtain
$(\partial_t \{h_i+h_e\},\varphi-h_i)=(\mbox{\boldmath$e$},
\mbox{\boldmath$j$})-(\mbox{\boldmath$e$},\mbox{\boldmath$curl$}\,\varphi)$.
Taking the Bean current-voltage relations into account we get
\[(\mbox{\boldmath$e$},\mbox{\boldmath$j$})=
(|\mbox{\boldmath$e$}|,|\mbox{\boldmath$j$}|)=(|\mbox{\boldmath$e$}|,j_c)\ge
(|\mbox{\boldmath$e$}|,|\mbox{\boldmath$curl$}\,\varphi|)\ge(\mbox{\boldmath$e$},
\mbox{\boldmath$curl$}\,\varphi)\] and arrive at the variational
inequality for the induced field $h_i$: \beq
\begin{array}{c}\mbox{Find}\ h_i\in K\ \mbox{such that}\ \
(\partial_t \{h_i+h_e\},\varphi-h_i)\ge 0\ \mbox{for any}\ \varphi\in K,\\
h_i(x,0)=h_0(x)-h_e(0).\end{array}\label{sc_vi}\eeq In the general
three-dimensional case the $h$-formulation of the Bean model can
also be derived \cite{Euro2} but then the magnetic field should be
determined in the whole space. A similar variational formulation
in terms of the current density we derive now is probably more
convenient.

Provided that no current is fed into a superconductor by electric
contacts,  $ {\nabla}\cdot\, \mbox{\boldmath$j$}=0 \ \mbox{in}\
\Omega, \ \ \mbox{\boldmath$j$}_n=0 \ \mbox{on}\
 \Gamma.
 $
Let us define the set of admissible current densities in $\Omega$,
\[ \bm{j}\in K=\left\{\bm{\varphi}(x,t)\,\left|\,\begin{array}{c}
\nabla\cdot\bm{\varphi}=0,\ \ |\bm{\varphi}|\le j_c\ \mbox{in}\
\Omega
\\\bm{\varphi}_n=0\ \mbox{on}\
\Gamma\end{array}\right.\right\},\]  express the electric field
via the vector and scalar magnetic potentials \cite{Jack},
\[\mbox{\boldmath$e$}+\partial_t
\mbox{\boldmath$A$}+\nabla\psi =0,\] and exclude the scalar
potential by multiplying this equation by $\bm{\varphi}-\bm{j}$
and integrating: $(\mbox{\boldmath$e$}+\partial_t
\mbox{\boldmath$A$},\mbox{\boldmath$\varphi$}-\mbox{\boldmath$j$})=0$
for any $\bm{\varphi}\in K$. Just as above, if $\bm{e}$ and $
\bm{j}$ are parallel and satisfy the Bean model relations
(\ref{BeanCond}), the inequality
$(\mbox{\boldmath$e$},\mbox{\boldmath$\varphi$}-\mbox{\boldmath$j$})\le
0$ holds for every admissible test function $\bm{\varphi}$. Hence,
$(\partial_t
\mbox{\boldmath$A$},\mbox{\boldmath$\varphi$}-\mbox{\boldmath$j$})\ge
0$.

Up to the gradient of a scalar function, determined by the gauge
and eliminated by scalar product with the test functions, the
vector potential is a convolution of the Green function of Laplace
equation, $G=\frac{1}{4\pi|x|}$, and the total current:
$$\bm{A}=G\ast \{\bm{j}+\bm{j}_e\}.$$ We arrived at the evolutionary
variational inequality with an ``implicit'' derivative in time:
\beq
\begin{array}{c}\mbox{Find}\ \bm{j}\in K\ \mbox{such that}\\
(G\ast \partial_t \{\bm{j}+\bm{j}_e\},\bm{\varphi}-\bm{j})\ge 0\
\mbox{for any}\ \bm{\varphi}\in K,\\
\bm{j}|_{t=0}=\bm{j}_0(x),\end{array}\label{Bean_vi}\eeq where
$\bm{j}_0=\nabla\times\bm{h}_0|_{\Omega}\in K$ is a given initial
current density distribution.

Experiments on hard superconductors are often performed on thin
flat samples, and we present also a scalar version of this
variational inequality for thin films in a perpendicular uniform
external magnetic field. We now assume that it is the sheet
current density, obtained by  integration of bulk current density
across the film thickness and also denoted
$\mbox{\boldmath$j$}(x,t)$, $x\in \Omega\subset \Rset^2$, obeys
the Bean's current-voltage relations. This current density should
also satisfy the conditions \beq {div}\, \mbox{\boldmath$j$}=0 \
\mbox{in}\ \Omega, \ \ \mbox{\boldmath$j$}_n=0 \ \mbox{on}\
\Gamma,\label{div0}\eeq where $div$ is the two-dimensional
divergence. For simplicity, we assume the domain $\Omega$ is
simply connected. Due to conditions (\ref{div0}) there exists a
stream function $h(x,t)$ such that
$\mbox{\boldmath$curl$}\,h=\mbox{\boldmath$j$}$ in $\Omega$ and
$h=0$ on ${\Gamma}$. Although $h$ is not the induced magnetic
field as in the case of a long cylinder in a parallel field, since
$|\bm{curl}\, h|=|\nabla h|$ this function belongs to the same set
$K$ of admissible functions (\ref{KK}). Let
$\mbox{\boldmath$j$}^{\prime}$ be another vector function
satisfying (\ref{div0}) and the condition
$|\mbox{\boldmath$j$}^{\prime}|\le j_c$ in $\Omega$, and
$\varphi\in K$ be the corresponding stream function. The external
vector potential $\bm{A}_e$, corresponding to the uniform
perpendicular magnetic field $h_e(t)$, can be chosen parallel to
the film; then ${curl}\, \bm{A}_e=h_e(t)$. Substituting into the
inequality (\ref{Bean_vi}) the curls of $h$ and $\varphi$ instead
of the current and test function, correspondingly, using the Green
theorem, and taking into account that
$\mbox{\boldmath$curl$}\,u\cdot\mbox{\boldmath$curl$}\,v=\nabla
u\cdot\nabla v$ we obtain a scalar variational inequality in terms
of the stream function: \beq
\begin{array}{c} \mbox{Find}\ h\in K\ \ \mbox{such that}\\
a(\partial_t h,\varphi - h)+(\partial_th_e,\varphi-h)\ge 0\
\mbox{for any}\ \varphi\in K,\\
h(x,0)=h_0(x),\end{array}\label{vi}\eeq where
$a(u,v)=\int_{\Omega}\int_{\Omega}\nabla u(x)\cdot\nabla
v(x')/\{4\pi |x-x'|\}\,dx\,dx'$.

The existence and uniqueness of solutions to (\ref{Bean_vi}) and
(\ref{vi}) were proved in \cite{Euro2} and \cite{BP}; see
\cite{JCP2} for the numerical solution of (\ref{vi}). It has been
shown in \cite{Euro2} that the effective resistivity $\rho$,
excluded in transition to the variational formulation
(\ref{sc_vi}), is a Lagrange multiplier related to the current
density constraint. Similar variational formulations may be
derived  for  much more general  current-voltage relations (see,
e.g., \cite{Boss,IEEE,Badia}) and present a very convenient
description of hysteretic magnetization typical of hard
superconductors. In particular, the critical current density $j_c$
depends usually on the magnetic field \cite{Kim}. Then the set of
admissible functions $K$ depends on the unknown solution and the
inequalities become quasivariational \cite{Euro2}. The power law
$|\bm{e}|=e_0(|\bm{j}|/j_c)^p$ is often employed instead of the
Bean's current voltage relation to account for the creep of
magnetic flux \cite{Brandt}; as $p\rightarrow\infty$, such model
converges to the Bean model \cite{BP,Yin}. Thermal fluctuations in
a superconductor may cause avalanches of magnetic vortices
resembling sand avalanches \cite{deJ}; in the Bean model these
avalanches correspond to discontinuous solutions of variational
inequality (\ref{Bean_vi}) with the jumps induced by local
fluctuations of the critical current density.

\subsection{Elastoplastic solids} The variational inequality
formulation for models in perfect elastoplasticity is well known
\cite{DL}. We briefly present this formulation to underline its
similarity to the variational formulations above.

Let an elastoplastic body occupy the domain $\Omega\subset
\Rset^3$ and the conditions of equilibrium,
\[\int_{\Omega}\mbox{\boldmath$g$}+\int_{\Gamma}\mbox{\boldmath$f$}=0,\ \ \ \
\int_{\Omega}\mbox{\boldmath$x$}\times\mbox{\boldmath$g$}+
\int_{\Gamma}\mbox{\boldmath$x$}\times\mbox{\boldmath$f$}=0,\]
hold for the given body force $\mbox{\boldmath$g$}$ and surface
traction $\mbox{\boldmath$f$}$. The stress tensor
$\mbox{\boldmath$\sigma$}$ should satisfy the local equilibrium
conditions (the usual summation convention is implied) \beq
\sigma_{ij,j}+g_i=0\ \mbox{in}\ \Omega, \ \ \ \
\sigma_{ij}n_j=f_i\ \ \mbox{on}\ \Gamma .\label{equ}\eeq Under the
assumption of small strain we have \beq
\epsilon_{ij}=\frac{1}{2}(u_{i,j}+u_{j,i}),\label{displ}\eeq where
$\mbox{\boldmath$u$}$ is the displacement vector field,
$u_{i,j}=\partial u_i/\partial x_j$.

It is assumed that the strain tensor can be presented as a sum of
elastic and plastic components, $
$\mbox{\boldmath$\epsilon$}$=$\mbox{\boldmath$e$}$+$\mbox{\boldmath$p$}$,$
where the elastic component obeys the linear Hooke's law,
$e_{ij}=A_{ijkl}\,\sigma_{kl}$. The plastic part is governed by an
incremental flow rule,
$\mbox{\boldmath$p$}=\int_0^t\dot{\mbox{\boldmath$p$}}dt$, in
which $\dot{\mbox{\boldmath$p$}}=\partial_t\mbox{\boldmath$p$}$ is
determined as follows. For a prescribed convex yield function
$\mathcal{F}(\mbox{\boldmath$\sigma$})$, it is postulated that the
stress tensor everywhere satisfies \beq
\mathcal{F}(\mbox{\boldmath$\sigma$})\le 0\label{F}\eeq and that
\[\dot{p}_{ij}=\lambda\,
\partial{\mathcal F}/\partial{\sigma_{ij}},\] where  $\lambda(x,t)\ge 0$
is the deformation rate such that $ {\mathcal
F}(\mbox{\boldmath$\sigma$}(x,t))< 0\longrightarrow
\lambda(x,t)=0.$ Let the admissible set $K$ be the set of  stress
tensors satisfying (\ref{equ}) and (\ref{F}) and
$\mbox{\boldmath$\tau$}\in K$. Using the strain-displacement
relation (\ref{displ}), the Green formula, and the equilibrium
conditions (\ref{equ}), one can show \cite{Prager} that
$(\dot{\epsilon}_{ij},\tau_{ij}-\sigma_{ij})=0$, hence \[
(A_{ijkl}\,\dot{\sigma}_{kl},\tau_{ij}-\sigma_{ij})+\left(\lambda\,
\partial{\mathcal
F}/\partial{\sigma_{ij}},\tau_{ij}-\sigma_{ij}\right)=0.\] The
second term here is nonpositive. Indeed, if ${\mathcal
F}(\mbox{\boldmath$\sigma$}(x,t))<0$ then $\lambda(x,t)=0$.
Otherwise  ${\mathcal F}(\mbox{\boldmath$\sigma$}(x,t))=0$ and,
because $\mathcal F$ is convex,
 ${\mathcal
F}(\{1-\theta\}\mbox{\boldmath$\sigma$}(x,t)
+\theta\mbox{\boldmath$\tau$}(x,t))\le 0$ for
any
$\theta\in[0,1]$. Therefore,
\[\left.\frac{d}{d\theta}{\mathcal
F}(\{1-\theta\}\mbox{\boldmath$\sigma$}+\theta\mbox{\boldmath$\tau$})
\right|_{\theta=+0}=\frac{\partial\mathcal
F(\mbox{\boldmath$\sigma$})}{\partial
\sigma_{ij}}(\tau_{ij}-\sigma_{ij})\le 0.\] We arrived at the
variational inequality:
\[\begin{array}{c}
\mbox{Find \boldmath$\sigma$}\in K\ \mbox{such that} \
(A_{ijkl}\,\dot{\sigma}_{kl},\tau_{ij}-\sigma_{ij})\ge 0\
\mbox{for any}\
\mbox{\boldmath$\tau$}\in K,\\
\mbox{\boldmath$\sigma$}(x,0)=\mbox{\boldmath$\sigma$}_0(x).
\end{array}\]
\section{Dual formulations for conjugate variables}
Although a common feature of the considered mathematical models is
the presence of conjugate variables, the variational inequalities
above are written for only one of them: the free surface in the
model for sandpile growth, current density in critical-state
superconductivity problems, stress tensor in elastoplasticity
problems. The dual variables, i.e, the surface flux, the electric
field, and the strain, correspondingly, have been eliminated in
transition to the variational inequalities. These inequalities can
be solved efficiently, however, knowledge of the primary variables
generally does not make determining the dual ones easy because the
constitutive relations are multivalued.

In elastoplasticity with hardening, a dual variational formulation
for strain has been derived and comprehensively studied by Han and
Reddy \cite{HR1}. Mathematically, this problem takes the form of
the so-called mixed variational inequality,
\[\begin{array}{c}\mbox{Find}\ u:[0,T]\rightarrow V\ \mbox{such
that for almost all}\ t\\
a(u(t),v-\partial_tu(t))-(f(t),v-\partial_tu(t))+\phi(v)-\phi(\partial_tu(t))\ge
0\
\mbox{for all}\ v\in V,\\
u(0)=u_0,\end{array}\] where $V$ is a Banach space, $a(.,.)$ is a
symmetric bilinear form, $f(t)\in V'$ is a linear functional, and
$\phi(.)$ is a convex, positively homogeneous, nonnegative
functional on $V$. This problem is not a standard variational
inequality although it resembles the parabolic variational
inequalities of the second type \cite{GLT}.

Strain formulation for the perfect plasticity problem considered
above turns out to be much more complicated because, without
hardening, the arising problem is not coercive in the usual
Sobolev spaces (coercive in a non-reflexive Banach space) and the
solution has to be sought in the space $BD(\Omega)$ of functions
of bounded deformation (\cite{Temam}, see \cite{EbobReddy} for a
review of recent results). Physically, the difference is
manifested in the ability of perfectly plastic materials to form
slip surfaces on which the tangential component of displacement is
discontinuous.

Similar difficulties arise in dual variational formulations of
other critical-state problems described in the first part of our
work and this, indeed, seems to be physically meaningful. Thus,
the continuity of only the normal component of surface sand flux
follows from the mass conservation law in the pile growth model.
Also, according to Maxwell equations, only the tangential
component of the electric field has to be continuous.
Mathematically, the corresponding problems are not coercive in
reflexive Banach spaces.

Below, we derive formally variational formulations in terms of the
dual variables for two critical-state problems where the primary
formulation is a variational (not quasivariational) inequality.
The questions of existence, uniqueness, and numerical
approximation of these problems need further investigation; we are
going to consider these questions in a separate publication
\cite{JWB2}.

\subsection{Surface flux in the model of pile growth}
Determining the flux of granular material pouring down the free
surface of a growing pile is necessary, e.g., if the material is
polydisperse and it is needed to predict the resulting
distribution of different species inside the pile (see
\cite{ChESc}). Let us assume the initial support $h_0$ in the pile
growth model has no steep slopes, $|\nabla h_0|\le k$ in $\Omega$.
In this case the model (\ref{m})-(\ref{bound}) can be written as
\beq\begin{array}{c}\partial_th+\nabla\cdot
\bm{q}=f,\\h|_{t=0}=h_0,\ \ \
\bm{q}_n|_{\Gamma}=0,\end{array}\label{hq}\eeq where the flux
$\bm{q}$ has the direction of $-\nabla h$ and the following
flux-slope relation holds:
\begin{equation} |\nabla h|\le k,\ \ |\nabla h|<
k\rightarrow \bm{q}=0.\label{Dcond}\end{equation} As it was shown
above, the free surface $h(x,t)$ can be sought as a solution of an
evolutionary variational inequality.

To derive a variational formulation of this model in terms of the
surface flux, let us define
$$V=H_0(div;\Omega)\triangleq\{\bm{\varphi}\in \bm{L}^2(\Omega)\ |\
\nabla\cdot\bm{\varphi}\in {L}^2(\Omega),\
\bm{\varphi}_n|_{\Gamma}=0\},$$ assume that $\bm{q}$ and $h$
satisfy the model relations (\ref{hq})-(\ref{Dcond}), and choose
an arbitrary test flux $\tilde{\bm{q}}\in V$. Using the
constitutive relations (\ref{Dcond}) we obtain
$$\nabla h\cdot(\tilde{\bm{q}}-\bm{q})\ge -|\nabla h||\tilde{\bm{q}}|-\nabla
h\cdot\bm{q}=-|\nabla h||\tilde{\bm{q}}|+k|{\bm{q}}|\ge
-k|\tilde{\bm{q}}|+k|{\bm{q}}|.$$ Hence,
$$(\nabla h,\tilde{\bm{q}}-\bm{q})\ge\phi(\bm{q})-\phi(\tilde{\bm{q}}),$$
where $\phi(\bm{q})=k\int_{\Omega}|{\bm{q}}|$. Since $(\nabla
h,\tilde{\bm{q}}-\bm{q})=-(h,\nabla\cdot\{\tilde{\bm{q}}-\bm{q}\})$,
we have
\[\phi(\tilde{\bm{q}})-\phi(\bm{q})-(h,\nabla\cdot\{\tilde{\bm{q}}-\bm{q}\})\ge 0.\] Let us
define $\bm{u}=\int_0^t\bm{q}\,dt.$ Then $\partial_t\bm{u}=\bm{q}$
and, from (\ref{hq}),  $\nabla\cdot \bm{u}=-h+h_0+\int_0^tf\,dt$.
We finally arrive at the following variational problem:
\begin{equation}
\begin{array}{c}\mbox{Find $\bm{u}:[0,T]\rightarrow V$
such that for any
$\tilde{\bm{q}} \in V$ and almost all $t$}\\
(\nabla\cdot\bm{u},\nabla \cdot \{ \tilde{\bm{q}}-
\partial_t\bm{u} \} )-( \mathcal{F},\nabla \cdot \{ \tilde{
\bm{q}}- \partial_t\bm{u} \}
)+\phi(\tilde{\bm{q}})-\phi(\partial_t\bm{u})\ge 0,\\
\mbox{and}\ \bm{u}|_{t=0}=0,
\end{array}\label{hqvi}\end{equation} where
$\mathcal{F}=h_0+\int_0^tf\,dt$. Since the problem is not coercive
in $V$ (coercive in a non reflexive Banach space), it may have and
may have no solution, which means an appropriate regularization is
needed. We do not investigate this issue further in a this work
and only note that, after discretization in time, the problem
becomes equivalent to a non-smooth optimization problem for each
time layer. Indeed, let $\bm{q}=(\bm{u}^{n+1}-\bm{u}^n)/\Delta t$
and $\bm{u}=\bm{u}^n+\frac{\Delta t}{2}\bm{q}$ be approximate
values at $t=\Delta t(n+\frac{1}{2})$. For each time layer we
obtain
\[
\phi(\tilde{\bm{q}})-\phi(\bm{q})+(\nabla\cdot
\{\bm{u}^n+\frac{\Delta t}{2}\bm{q}\},\nabla \cdot
\{\tilde{\bm{q}}- \bm{q}\})-(\mathcal{F}^{n+1/2},\nabla \cdot
\{\tilde{\bm{q}}-\bm{q}\})\ge 0,
\]
which is equivalent to
\begin{equation}\begin{array}{lcl}\bm{q}^{n+\frac{1}{2}}=\arg
&\min &\left\{\frac{\Delta t}{4} (\nabla\cdot \bm{q},\nabla\cdot
\bm{q})+\phi(\bm{q})+(\nabla \cdot
\bm{u}^n-\mathcal{F}^{n+1/2},\nabla \cdot \bm{q})\right\}\\
&\bm{q}\in V&\
\end{array}\label{optS}\end{equation}
Provided the latter problem has a solution, approximate value of
$\bm{u}$ on the next time layer can be found as
$\bm{u}^{n+1}=\bm{u}^{n}+\Delta t\bm{q}^{n+\frac{1}{2}}$.

\subsection{Electric field in the Bean model for superconductors}
The $h$- and $j$-variational formulations of the Bean model can be
used for efficient computation of magnetic fields, current
densities and, therefore, forces and their moments in various
applications of type-II superconductors. 
However, even if the current density is known, the electric field
is not determined in a unique way by the assumption that the
directions of $\bm{e}$ and $\bm{j}$ coincide and the
current-voltage relation
\begin{equation}|\bm{j}|\le j_c,\ \ |\bm{j}|< j_c\rightarrow\
\bm{e}=0\label{EJ}\end{equation} or another constitutive law
described by a monotone multivalued graph is satisfied. That is
why calculating the electric field in a superconductor is
generally a non-trivial task.\footnote{The case of an infinite
cylinder in a perpendicular external magnetic field is an
exception.} In particular, this field is necessary to estimate the
local AC loss $\bm{e}\cdot \bm{j}=j_c|\bm{e}|$ that causes heating
and thermal instability of superconductors.

For some simple configurations, the electric field in
superconductors has been considered in \cite{BrandtE,BrandtEF}
and, recently, for a generalized Bean model in \cite{BmL}. Here we
propose a completely different approach based not on the
determination of the magnetic field and subsequent integration of
Faraday's law along the flux penetration streamlines but on a
direct variational reformulation of the Bean model in terms of
electric field. 

Let $W=H(curl;\Omega)\triangleq\{\bm{\varphi}\in
\bm{L}^2(\Omega)\,|\,\nabla\times \bm{\varphi}\in
\bm{L}^2(\Omega)\}$ be the space of electric fields in the
superconductive domain $\Omega\subset \mathbb{R}^3$. Assuming
directions of $\bm{j}$ and $\bm{e}$ coincide and the Bean model
relations are satisfied in this domain, for any test field
$\widetilde{\bm{e}}\in W$ we obtain
$$(\nabla\times
\bm{h})\cdot(\widetilde{\bm{e}}-\bm{e})=\bm{j}\cdot(\widetilde{\bm{e}}-\bm{e})\le
j_c|\widetilde{\bm{e}}|-
\bm{j}\cdot\bm{e}=j_c|\widetilde{\bm{e}}|-j_c|\bm{e}|.$$
Integrating over $\Omega$ we get $(\nabla\times
\bm{h},\widetilde{\bm{e}}-\bm{e})\le\phi(\widetilde{\bm{e}})-\phi(\bm{e}),$
where $\phi(\bm{e})=j_c\int_{\Omega}|\bm{e}|$. On the other hand,
$$(\nabla\times
\bm{h},\widetilde{\bm{e}}-\bm{e})=(\bm{h},\nabla\times\{\
\widetilde{\bm{e}}-\bm{e}\})+\int_{\Gamma}(\bm{h}\times \{
\widetilde{\bm{e}}-\bm{e}\})\cdot \bm{n}.$$ Let
$\bm{u}=-\bm{A}_0+\int_0^t\bm{e}dt$, where the vector potential
$\bm{A}_0=G\ast (\bm{j}+\bm{j}_e)|_{t=0}$ satisfies
$\bm{h}_{0}=\nabla\times \bm{A}_0$. Then
$$\partial_t\bm{u}=\bm{e}, \ \ \ \ \nabla\times
\bm{u}=-\bm{h}_0+\int_0^t\nabla\times
\bm{e}\,dt=-\bm{h}_0-\int_0^t\partial_t\bm{h}dt=-\bm{h}.$$  
 The following inequality is thus
obtained: for any $\widetilde{\bm{e}}\in W$
\begin{equation}(\nabla\times
\bm{u},\nabla\times\{ \widetilde{\bm{e}}-\partial_t\bm{u} \})-
\int_{\Gamma}(\bm{h}\times \{ \widetilde{\bm{e}}-\partial_t\bm{u}
\} )\cdot \bm{n} +\phi(\widetilde{\bm{e}})-\phi(\partial_t \bm{u})
\ge 0.\label{notVF}\end{equation} In the general case this is  not
yet the needed ${e}$-formulation because it contains  the
tangential component of magnetic field on $\Gamma$. However, for
an infinite cylinder in a parallel uniform external magnetic field
$\bm{h}|_{\Gamma}=\bm{h}_e(t)$, where $\bm{h}_e$ is the field
generated by the external current, so
\begin{equation}(\nabla\times
\bm{u},\nabla\times\{\widetilde{\bm{e}}-\partial_t\bm{u}\})-\bm{h}_e(t)\cdot\int_{\Gamma}\bm{n}
\times(\widetilde{\bm{e}}-\partial_t\bm{u})+\phi(\widetilde{\bm{e}})-\phi(\partial_t\bm{u})\ge
0\label{cyl}\end{equation} and it is not difficult to see that,
after discretization in time, (\ref{cyl}) becomes equivalent to a
non-smooth optimization problem similar to (\ref{optS}).

Let us now return to the general case and consider an auxiliary
boundary value problem in the exterior domain
$\omega$,\footnote{Since the displacement current is omitted, in
our model the electric field in an insulator (the outer space) is
not unique. The magnetic field, however, is.}
\begin{equation}\begin{array}{c}\partial_t\bm{h}+\nabla\times \bm{e}=0,\
\ \ \nabla\times \bm{h}=\bm{j}_e,\\ \bm{h}|_{t=0}=\bm{h}_0,\ \ \
\bm{e}_{\tau}|_{\Gamma}=\bm{\mathcal{E}}
,\end{array}\label{bvp}\end{equation} where
$\bm{e}_{\tau}=\bm{n}\times(\bm{e}\times \bm{n})|_{\Gamma}$ and
$\bm{\mathcal{E}}$ is a tangential field given on $\Gamma$. We
choose the vector potential of external current as $\bm{A}_e=G\ast
\bm{j}_e$ and define the field
$\bm{h}_e=\nabla\times(\bm{A}_e-\bm{A}_e|_{t=0})+\bm{h}_0$. In
$\omega$ this field satisfies
\[\nabla\times \bm{h}_e=\bm{j}_e,\ \ \nabla\cdot \bm{h}_e=0,\ \ \bm{h}_e|_{t=0}=\bm{h}_0.\]
 Let us set $\bm{h}=\bm{h}_e+\bm{H}$,
$\bm{e}=-\partial_t\bm{A}_e+\bm{E}$ and rewrite problem
(\ref{bvp}) as
\begin{equation}\begin{array}{c}\partial_t\bm{H}+\nabla\times \bm{E}=0,\
\ \ \nabla\times \bm{H}=0,\\ \bm{H}|_{t=0}=0,\ \ \
\bm{E}_{\tau}|_{\Gamma}=\bm{\mathcal{E}}_1,\end{array}\label{bvp1}\end{equation}
where
$\bm{\mathcal{E}}_1=\bm{\mathcal{E}}+\partial_t\bm{A}_{e,\tau}$.
Defining $ \bm{U}=\int_0^t\bm{E}dt$ and integrating in time the
equations from (\ref{bvp1}) containing $\bm{E}$  yield
\begin{equation}\begin{array}{c}\bm{H}+\nabla\times \bm{U}=0,\ \ \
\nabla\times \bm{H}=0,\\
\bm{U}_{\tau}|_{\Gamma}=\bm{\mathcal{U}}
,\end{array}\label{bvp3}\end{equation} where
$\bm{\mathcal{U}}=\int_0^t\bm{\mathcal{E}}dt+\bm{A}_{e,\tau}(x,t)-\bm{A}_{e,\tau}(x,0)$.

Let us note that we also have $\nabla\cdot \bm{H}=0$ a.e. in
$\omega$, so it makes sense to seek a solution $\bm{H}$ in the
Sobolev space $\bm{H}^1(\omega)$. Denote $X=\{\bm{\psi}\in
\bm{H}^1(\omega)\ |\ \nabla\times\bm{\psi}=0 \}$. The problem
(\ref{bvp3}) admits a weak formulation,
\begin{equation}
\mbox{Find}\ \bm{H}\in X\ \mbox{such that}\
(\bm{H},\bm{\psi})=\int_{\Gamma}(\bm{\mathcal{U}}\times\bm{\psi})\cdot
\bm{n},\ \ \forall \bm{\psi}\in X,\label{varpr}\end{equation}
where the normal $\bm{n}$ is directed inside $\omega$ and the
integral on the right is understood as the duality pairing on
$H^{-1/2}\times H^{1/2}$. Existence of a unique solution to this
problem follows from the Lax-Milgram theorem. This defines a
linear operator $K: \bm{\mathcal{U}}\rightarrow
\bm{H}_{\tau}|_{\Gamma}$ acting from $H^{-1/2}(\Gamma)$ to
$H^{1/2}(\Gamma)$. It is not difficult to see that this operator
is symmetric in the following sense:  for any two functions
$\bm{v},\,\bm{w}\in H^{-1/2}(\Gamma)$ there holds
\begin{equation}\int_{\Gamma}(K(\bm{v})\times \bm{w})\cdot \bm{n}=\int_{\Gamma}(K(\bm{w})\times \bm{v})\cdot
\bm{n}.\label{eqeq}\end{equation} Indeed, let $\bm{H}^v$ and
$\bm{H}^w$ be the solutions of (\ref{varpr}) with
$\bm{\mathcal{U}}=\bm{v}$ and $\bm{\mathcal{U}}=\bm{w}$,
correspondingly. Then
\begin{eqnarray*}(\bm{H}^v,\bm{H}^w)=\int_{\Gamma}(\bm{v}\times \bm{H}^w)\cdot
\bm{n}=\int_{\Gamma}(\bm{v}\times K(\bm{w}))\cdot \bm{n},\\
(\bm{H}^w,\bm{H}^v)=\int_{\Gamma}(\bm{w}\times \bm{H}^v)\cdot
\bm{n}=\int_{\Gamma}(\bm{w}\times K(\bm{v}))\cdot
\bm{n}\end{eqnarray*} and (\ref{eqeq}) is proved. We also see that
$\int_{\Gamma}(\bm{v}\times K(\bm{v}))\cdot
\bm{n}=(\bm{H}^v,\bm{H}^v)\ge 0$ for any $\bm{v}\in
H^{-1/2}(\Gamma)$.

Let us now choose $\bm{\mathcal{E}}$ to be the continuous
tangential component of the electric field on $\Gamma$,
$\bm{\mathcal{E}}=\bm{e}_{\tau}$. Then
\begin{align*}\bm{\mathcal{U}}&=\int_0^t{\bm{e}_{\tau}}dt
+\bm{A}_{e,\tau}(x,t)-\bm{A}_{e,\tau}(x,0)
=\bm{u}_{\tau}+\bm{A}_{0,\tau}+\bm{A}_{e,\tau}-\bm{A}_{e,\tau}|_{t=0}\\
\ & =\bm{u}_{\tau}+G\ast (\bm{j}|_{t=0}+\bm{j}_e)\end{align*} and
the tangential component of the magnetic field can be presented as
\[\bm{h}_{\tau}=\bm{h}_{e,\tau}+\bm{H}_{\tau}=
K(\bm{u}_{\tau})+\bm{\mathcal{F}},\] where
$\bm{\mathcal{F}}=\bm{h}_{e,\tau}+K(\{G\ast
(\bm{j}|_{t=0}+\bm{j}_e)\}_{\tau})$.

We can now rewrite (\ref{notVF}) as a mixed variational
inequality:
\begin{equation}\begin{array}{c}
\mbox{Find $\bm{u}:[0,T]\rightarrow W$ such that for any
$\widetilde{\bm{e}} \in W$ and almost all $t$}
\\
(\nabla\times
\bm{u},\nabla\times\{\widetilde{\bm{e}}-\partial_t\bm{u}\})-\int_{\Gamma}(K(\bm{u}_{\tau})
\times\{\widetilde{\bm{e}}-\partial_t\bm{u}\})\cdot\bm{n}-\\
\int_{\Gamma}(\bm{\mathcal{F}}
\times\{\widetilde{\bm{e}}-\partial_t\bm{u}\})\cdot
\bm{n}+\phi(\widetilde{\bm{e}})-\phi(\partial_t\bm{u})\ge 0\\
\mbox{and}\ \bm{u}|_{t=0}=-\bm{A}_0.
\end{array}\label{VF}\end{equation} Discretizing in time as
above and making use of the equality (\ref{eqeq}) we obtain, for
each time layer, a non-smooth optimization problem:
\[\begin{array}{lcl}\bm{e}^{n+1/2}=\arg &\min& \Phi(\bm{e})
\\&\bm{e}\in W&
\end{array}\]
where
\begin{align*}
\Phi(\bm{e})= & \frac{\Delta t}{4}(\nabla\times
\bm{e},\nabla\times \bm{e})+\frac{\Delta
t}{4}\int_{\Gamma}(\bm{e}_{\tau} \times K(\bm{e}_{\tau}))\cdot
\bm{n}+\phi(\bm{e})
\\&-{\Delta
t}\int_{\Gamma}(\{K(\bm{e}^{n}_{\tau})+\bm{\mathcal{F}}^{n+1/2}\}\times
\bm{e}_{\tau})\cdot \bm{n} +(\nabla\times \bm{u}^n,\nabla\times
\bm{e})\end{align*} and $\bm{u}^{n+1}=\bm{u}^n+\Delta t
\bm{e}^{n+1/2}$.

\section*{Conclusion} To derive quasistationary critical-state models of
the spatially extended dissipative systems considered above, we
needed to specify only the possible direction of system's
evolution, and which changes of external conditions make the state
of the system unstable. The rate with which such a system driven
by the external forces rearranges itself is determined implicitly
by some conservation law coupled with a condition of equilibrium.
This rate appears in the model as a Lagrange multiplier related to
the equilibrium constraint. Although the conservation laws and
conditions of equilibrium may vary, the multiplicity of metastable
states, typical of many dissipative systems, is usually a
consequence of a unilateral constraint. This makes variational
inequalities a suitable tool for modeling these  systems.

Typically, some of the physically relevant conjugate variables are
eliminated in transition to the variational formulation of
critical-state problems. It is, however, possible to derive dual
formulations (mixed variational inequalities) in terms of these
variables. Although arising mathematical problems  need further
investigation, we believe these dual formulations will also serve
a basis for efficient numerical simulations.

 \end{document}